\documentclass[twocolumn,american]{revtex4-2}
\usepackage[T1]{fontenc}
\usepackage[utf8]{inputenc}
\usepackage{color}
\usepackage{babel}
\usepackage{array}
\usepackage{multirow}
\usepackage{amstext}
\usepackage{amssymb}
\usepackage{graphicx}
\usepackage[unicode=true,pdfusetitle,
 bookmarks=true,bookmarksnumbered=false,bookmarksopen=false,
 breaklinks=false,pdfborder={0 0 0},pdfborderstyle={},backref=false,colorlinks=true]
 {hyperref}
\hypersetup{
 urlcolor=blue, citecolor=blue}

\makeatletter

\newcommand{\lyxmathsym}[1]{\ifmmode\begingroup\def\b@ld{bold}
  \text{\ifx\math@version\b@ld\bfseries\fi#1}\endgroup\else#1\fi}

\providecommand{\tabularnewline}{\\}

\makeatother

\begin{document}
\title{Vibrational and structural properties of the $R$Fe$_{4}$Sb$_{12}$
($R=$Na, K, Ca, Sr, Ba) filled skutterudites}
\author{Juliana G. de Abrantes$^{1}$, Marli R. Cantarino$^{1}$, Wagner R.
da Silva Neto$^{1,2}$, Victória V. Freire$^{1}$, Alvaro G. Figueiredo$^{1}$,
Tarsis M. Germano$^{1}$, Bassim Mounssef Jr.$^{2}$, Eduardo M. Bittar$^{3}$,
Andreas Leithe-Jasper$^{4}$, Fernando A. Garcia$^{1}$}
\affiliation{$^{1}$Intituto de Física, Universidade de São Paulo, 05508-090, São
Paulo-SP, Brazil}
\affiliation{$^{2}$Instituto de Química, Universidade de São Paulo, 05508-090,
São Paulo-SP, Brazil}
\affiliation{$^{3}$Centro Brasileiro de Pesquisas Físicas, Rio de Janeiro, RJ
22290-180, Brazil}
\affiliation{$^{4}$Max Planck Institute for Chemical Physics of Solids, D-01187
Dresden, Germany.}
\begin{abstract}
Vibrational and elastic properties of the $R$Fe$_{4}$Sb$_{12}$
skutterudites are investigated by, respectively, temperature $(T)$
dependent extended X-ray absorption fine structure (EXAFS) and pressure
$(P)$ dependent x-ray diffraction (XRD) experiments. The Fe $K$-edge
EXAFS experiments of the $R=$ K, Ca and Ba materials were performed
in the $T$-interval $6<T<300$ K and XRD experiments of the $R=$
Na, K, Ca, Sr and Ba materials were performed in the $P$-interval
$1\text{ atm }<P<16$ GPa. From EXAFS, we obtained the correlated
Debye-Waller parameters that were thus analyzed to extract effective
spring constants connected with the Fe-$Y$ (where $Y=$ either $R$,
Fe or Sb) scattering paths. Our findings suggest that in the case
of the light cations, $R=$ K or Ca, the $R$ atoms are relatively
weakly coupled to the cage, in a scenario reminiscent to the Einstein
oscillators. From the XRD experiments, we obtained the bulk modulus
$B_{0}$ for all $R=$Na, K, Ca, Sr and Ba materials, with values
ranging from $77$ GPa ($R=$ K) to $R=99$ GPa ($R=$ Ba) as well
as the compressibility $\beta$ as a function of $P$. The trend in
$\beta$ as a function of the $R$ filler is discussed and it is shown
that it does not correlate with simple geometrical considerations
but rather with the filler-cage bonding properties.
\end{abstract}
\maketitle

\section{Introduction}

Filled skutterudites are materials with chemical formula $RT_{4}X_{12}$
where the $T_{4}X_{12}$ elements form a relatively rigid cage framework
inside which the $R$ elements are allocated \citep{jeitschko_lafe4p12_1977}.
In the simplest chemical bonding scenario, the cage is structured
by strong $T-X$ covalent bondings and the $R$ fillers generically
display a cationic character, donating electrons that further stabilize
the $T_{4}X_{12}$ cage. While it is well known that the description
of bonding in filled skutterudites is more involved \citep{gumeniuk_filled_2010},
this simple picture can be applied as a guide to discover new filled
skutterudites \citep{luo_large_2015} as well as to justify their
unusual vibrational dynamics \citep{keppens_localized_1998}.

Concerning the latter topic, research on the filled skutterudites
is indeed triggered mostly by the presence of an unusual vibrational
dynamics that makes the filled skutterudites good candidates for thermoelectric
applications \citep{snyder_complex_2008,mao_advances_2018,el_oualid_high_2021}.
In this regard, earlier contributions proposed that the $R$ fillers
behave as independent, non dispersive and low energy oscillators (or
rattlers), that scatter the cage derived phonons, giving rise to a
phonon glass, which impedes the thermal conduction and raises the
material thermoelectric figure of merit $ZT$ \citep{hermann_einstein_2005}.

On one account, the phonon glass scenario provided a first approach
to understand the skutterudites' vibrational dynamics. For instance,
filling the cages with distinct fillers may cause the smaller $R$
cations to display an unusual rattling behavior in the oversized cage,
as observed \citep{garcia_coexisting_2009,garcia_spin_2012}. This
scenario, however, was soon challenged \citep{koza_breakdown_2008}
and it is now well understood that the separation between rattler
and cage vibrational dynamics does not offer an adequate description
of the filled skutterudite vibrational properties \citep{wei_filling-fraction_2017,zhao_multi-localization_2015,gainza_unveiling_2020}.
All this understanding is particularly true in the case of the $R$Fe$_{4}$Sb$_{12}$
($R=$ Na, K, Ca, Sr, Ba, La and Yb) skutterudites \citep{schnelle_magnetic_2008,koza_vibrational_2010,koza_vibrational_2011,feldman_lattice-dynamical_2014,moechel_lattice_2011,koza_low-energy_2015}
and provides support for the design of more efficient antimony based
skutterudite materials \citep{prado-gonjal_extra-low_2017,bae_effective_2019,wang_filling_2017}.

On top of the good potential for thermoelectric applications, filled
skutterudites are very flexible platforms, displaying a broad range
of electronic, magnetic and thermal properties \citep{maple_new_2009}.
The $R$Fe$_{4}$Sb$_{12}$ ($R=$ Na, K, Ca, Sr, Ba and La) materials,
in particular, are itinerant magnets \citep{leithe-jasper_ferromagnetic_2003,leithe-jasper_weak_2004,berardan_rare_2005,matsuoka_nearly_2005,schnelle_magnetic_2008,leithe-jasper_neutron_2014}
whose magnetic properties are dominated by a high density of states
due to the mixing between the Fe derived $3d$ states and Sb derived
$5p$ states forming heavy bands in the vicinity of the Fermi level
\citep{kimura_infrared_2007,mounssef_hard_2019}.

In this work, we focus on the vibrational and structural properties
of the $R$Fe$_{4}$Sb$_{12}$ ($R=$ Na, K, Ca, Sr and Ba) filled
skutterudites as investigated, respectively, by temperature $(T)$
dependent extended x-ray absorption fine structure (EXAFS) and pressure
($P$) dependent x-ray diffraction (XRD) experiments. EXAFS was soon
recognized by Cao \emph{et al }\citep{cao_evidence_2004} as a suitable
probe for the site-specific vibrational dynamics in skutterudites.
Later contributions \citep{bridges_complex_2015,bridges_local_2016,keiber_modeling_2015}
developed the application of the technique to the correlated nature
of the filler-cage vibrational dynamics. Here, we focus on the Fe
$K$-edge EXAFS of the $R=$ K, Ca and Ba samples. In our investigation,
we find that the cage vibrational dynamics clearly depends on the
filler type, corroborating the correlated nature of the filler-cage
dynamics.

Our XRD experiments were carried out for pressures typically up to
$\approx16$ GPa for all samples, except in the case of BaFe$_{4}$Sb$_{12}$,
which was investigated up to $24$ GPa. Overall, the $P$-dependence
of all samples is well described by the Birch-Murnaghan model \citep{birch_finite_1947}
from which we extract the bulk modulus $B_{0}$ for each $R$. The
$B_{0}$ behavior as a function of $R$ is discussed in the context
of bonding specific properties of the $R$Fe$_{4}$Sb$_{12}$ skutterudites.
This discussion is supported by quantum chemistry calculations. Moreover,
starting at pressures about $\approx12$ GPa, the XRD results display
broad reflections. This finding suggests that the $R$Fe$_{4}$Sb$_{12}$
materials undergo a process of increasing lattice disorder triggered
by pressure, which is reminiscent of the pressure-induced partial
amorphization observed for other cage systems \citep{mardegan_pressure-induced_2013}.

\section{Methods}

High-quality polycrystalline samples of $R$Fe$_{4}$Sb$_{12}$ ($R=$Na,
K, Ca, Sr, and Ba) skutterudites were synthesized by a solid-state
sintering method \citep{leithe-jasper_ferromagnetic_2003}. Fe $K$-edge
EXAFS experiments of the $R=$ K, Ca, and Ba samples were performed
at the XAFS2 \citep{figueroa_upgrades_2016} beamline of the Brazilian
Synchrotron Light Source (CNPEM-LNLS). The spectra of the K and Ca
filled materials were measured by both fluorescence and transmission
modes, whereas the Ba filled material was measured only in the fluorescence
mode. A complete temperature scan in the interval $300>T>6$ K was
obtained for the three samples. An Fe foil, kept at room temperature,
was measured in the transmission mode as a reference throughout the
experiments. A conventional He-flow cryostat was employed to achieve
temperatures down to $T=6$ K.

The EXAFS data were analyzed by multiple scattering theory implemented
by the FEFF code \citep{rehr_theoretical_2000}. The graphical Demeter
platform \citep{ravel_athena_2005} was adopted to perform the fittings
of a structural model including single and multiple scattering paths
up to $5$ $\lyxmathsym{\AA}$ from the Fe absorber. To decrease the
number of parameters in the fitting, the EXAFS $S_{0}^{2}$ parameter
(the path amplitudes) was obtained from \emph{ab initio} calculations
implemented by the FEFF$8.4$ code. The\emph{ ab initio} calculations
were performed for clusters of $226$ atoms, adopting the Hedin-Lundqvist
pseudopotential to account for the exchange interaction \citep{mounssef_hard_2019}.

The powder XRD experiments were performed at the XDS \citep{lima_xds_2016}
beamline of the CNPEM-LNLS at room temperature. A diamond anvil cell
(DAC) was employed during experiments to achieve pressures up to $24$
GPa ($R=$ Ba), but typically up to $16$ GPa ($R=$ Na, K, Ca, and
Sr). A mixture of methanol and ethanol (in a $1:4$ proportion) was
adopted as a pressure media. The pressure was measured by monitoring
the Raman fluorescence from a ruby standard mounted in the DAC together
with the samples. In all measurements, the X-ray energy was set to
$20$ keV. The XRD powder profile fitting and background removal were
performed with the GSAS-II software \citep{toby_gsas-ii_2013}.

The filler-cage bonding properties of the $R=$ Ca-, Sr- and Ba-filled
materials were calculated within the density functional theory (DFT)
approximation, adopting the ORCA 5.0 program package \citep{Neese2020,Neese2012a}.
A supercell containing a total of $339$ atoms was generated for the
calculations which were implemented\emph{ }in\emph{ }an\emph{ embedded
approach}, shown to predict local properties with good accuracy \citep{Dittmer2019}.
Here, a cell of $45$ atoms containing one filler plus the cage atoms
was treated at a quantum level. This cell was segregated by $2$ atomic
layers of atoms described as capped effective core potentials (cECPs)
to avoid spurious electron leakage \citep{Dittmer2019}. The remaining
atoms were simulated in a molecular mechanics approach, with their
charges ascribed in agreement with the Zintl phase concept to constrain
the $24$ electron rule per $\mathrm{FeSb_{3}}$ unit \citep{luo_large_2015}
by means of a force field. All atomic positions were obtained from
crystallographic data. For all atoms, the BP$86$ GGA functional was
used along with the Karlsruhe valence triple $\zeta$ with one set
of polarization functions. Spin-orbit coupling effects were considered
but no sizable effects were observed in the constitution of most relevant
bonding MOs.

\section{Results and discussion}

\subsection{EXAFS analysis and vibrational properties}

\begin{figure*}
\begin{centering}
\includegraphics[width=1\textwidth]{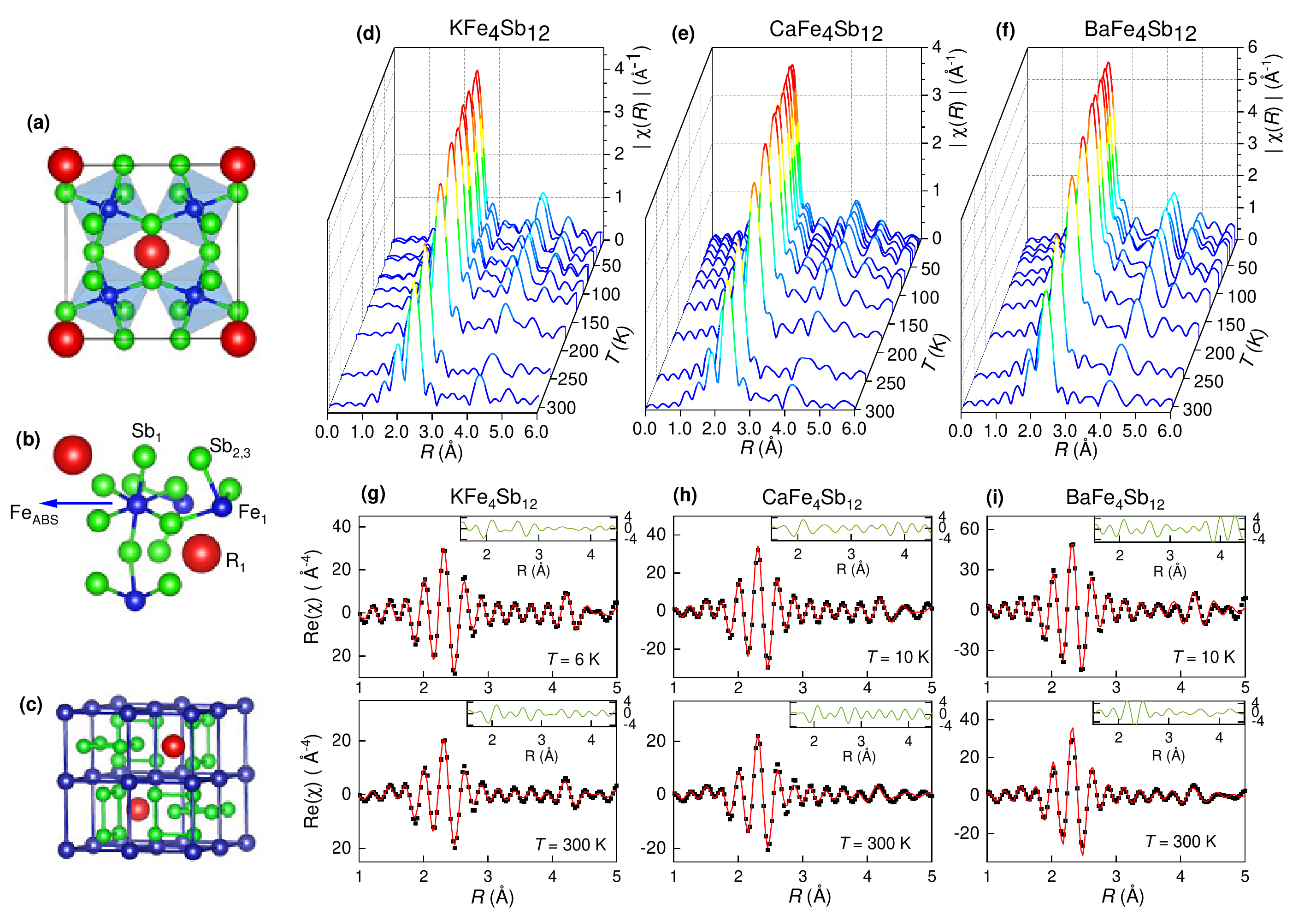}
\par\end{centering}
\caption{$(a)-(c)$ Three representations of the ternary filled skutterudite
structure (space group $Im\bar{3}$): $(a)$ the $R$Fe$_{4}$Sb$_{12}$
conventional unit cell highlighting the FeSb$_{3}$ distorted octahedra;
$(b)$ the structural elements considered in the EXAFS analysis; $(c)$
the vibrating elements of the 1D correlated rattling model (see text)
\citep{momma_vesta_2008}. In all cases, the $R$, Fe and Sb atoms
are represented by, respectively, red, blue and green spheres. $(d)-(f)$
Fourier transformed EXAFS spectra of the $R=$ K, Ca and Ba materials
in $T$-interval $6\text{ K}<T<300$ K. $(g)-(i)$ Low-$T$ (upper
panel) and high-$T$ (lower panel) representative EXAFS data and data
analysis (thick red line) for the $R=$ K, Ca and Ba materials. In
each panel, the inset shows the residuals of the presented fittings.
\label{fig:structure_exafsresults}}
\end{figure*}

Our EXAFS experiments focus on the Fe $K$-edge EXAFS of the $R$Fe$_{4}$Sb$_{12}$
($R=$ K, Ca and Ba) materials to understand how their vibrational
properties evolve as a function of the filler. Among the $R=$ Na,
K, Ca, Sr, and Ba filled materials, the changing periodic properties
concern the filler charge and the filler-cage mass/size relation.
By selecting the K and Ca filled materials, we probe the effects of
the filler charge, whereas the filler-cage mass relation is kept nearly
constant. Choosing the Ca and Ba filled cases, the filler-cage mass/size
relation is probed, whereas keeping the filler charge constant. Our
choice of $R$ thus suffices to capture the important trends of the
vibrational dynamics of the $R$Fe$_{4}$Sb$_{12}$ materials.

In figures \ref{fig:structure_exafsresults}$(a)-(c)$ we present
three representations of the skutterudite structure. The $R$ fillers,
Fe atoms and Sb atoms are represented, respectively, by red, blue
and green spheres. In figure \ref{fig:structure_exafsresults}$(a)$,
we show the typical conventional unit cell representation, where we
highlight the FeSb$_{6}$ octahedra. In figure \ref{fig:structure_exafsresults}$(b)$,
we centered the representation in a given Fe atom, which we call the
absorber, and leave only the atoms that are about $5$ $\lyxmathsym{\AA}$
apart from this Fe atom. In figure \ref{fig:structure_exafsresults}$(c)$,
we show an alternative representation, where we show the filler, the
square Sb ring and the Fe cage. This representation highlights the
vibrating components of the effective 1D model for the correlated
rattling dynamics in filled skutterudites \citep{keiber_modeling_2015,bridges_local_2016}
which we will discuss below.

The Fourier transformed data for all $T$ values are presented in
figures \ref{fig:structure_exafsresults}$(d)$, $(e)$ and $(f)$
for the $R=$ K, Ca and Ba materials, respectively. Our EXAFS analysis
is based upon single and multiple scattering paths within the cluster
depicted in figure \ref{fig:structure_exafsresults}$(b)$. This cluster
includes all atoms within $5$ $\lyxmathsym{\AA}$ from the absorber.
In this range, there are five Fe - $Y$ single scattering paths, where
$Y$ is the scattering element. The second and third Sb nearest neighbors,
denoted Sb$_{2}$ and Sb$_{3}$, are too close to each other, and
thus we adopt a single correlated Debye-Waller parameter $\sigma_{\text{Fe}-\text{Sb}_{2,3}}^{2}$
to describe both paths. In total, four $\sigma_{\text{Fe}-Y}^{2}$
are considered: $\sigma_{\text{Fe}-\text{Sb}_{1}}^{2}$, $\sigma_{\text{Fe}-\text{R}_{1}}^{2}$,
$\sigma_{\text{Fe}-\text{Sb}_{2,3}}^{2}$ and $\sigma_{\text{Fe}-\text{Fe}_{1}}^{2}$.
Single scattering paths dominate the EXAFS signal, but multiple scattering
paths were included to improve the data refinement. The disorder parameters
of the latter were modeled in terms of the single scattering path
$\sigma_{\text{Fe}-Y}^{2}$ parameters. In its turn, the $T$-dependence
of the $\sigma_{\text{Fe}-Y}^{2}$ parameters was modeled on the basis
of the Debye model for correlated disorder for all Fe-$Y$ paths but
the Fe-$R$ paths, which were analyzed under the Einstein model, as
typical for filled skutterudite materials \citep{cao_evidence_2004,bridges_complex_2015}.
Higher-order cumulants, which suggest the presence of anharmonic vibrations
\citep{hu_anharmonicity_2021}, were not adopted in our models. Representative
fittings and residuals (for low and high $T$ data) for the $R=$
K, Ca and Ba cases are displayed in figures \ref{fig:structure_exafsresults}$(g)-(i)$,
showing that the structural model adopted in this paper offers a fair
description of the data in all $T$-interval.

\begin{figure}
\begin{centering}
\includegraphics[width=1\columnwidth]{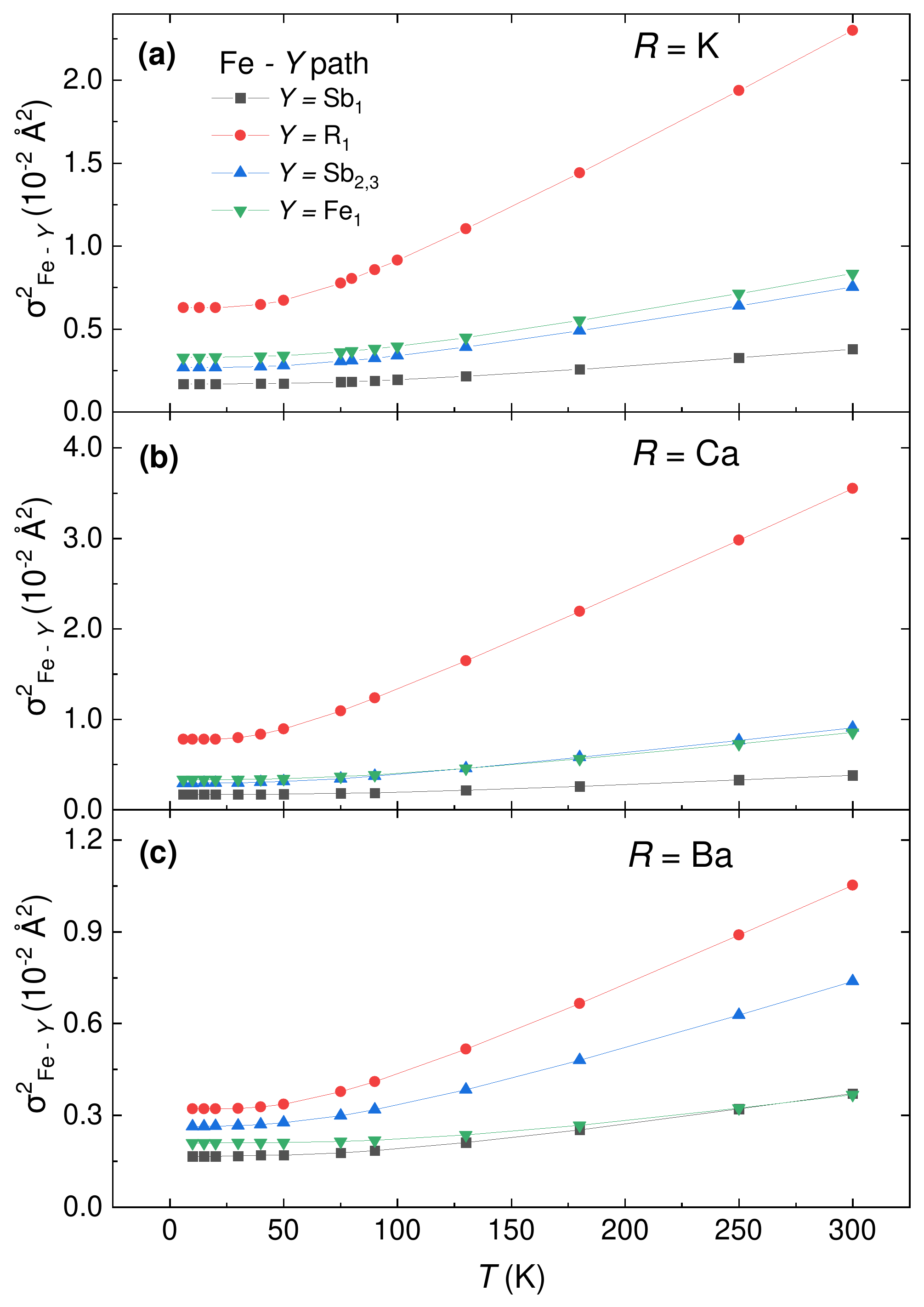}
\par\end{centering}
\caption{Temperature dependence of the correlated Debye-Waller parameters obtained
from the Fe $K$-edge EXAFS analysis for the $(a)$ KFe$_{4}$Sb$_{12}$,
$(b)$ CaFe$_{4}$Sb$_{12}$ and $(c)$ BaFe$_{4}$Sb$_{12}$ samples.
\label{fig:exafsparameters}}
\end{figure}

In figures \ref{fig:exafsparameters}$(a)-(c)$ we show the correlated
Debye-Waller parameters obtained from the EXAFS experiments. Whereas
the Fe-Sb$_{1}$ and Fe-Sb$_{2,3}$ bond disorder parameters are nearly
$R$-independent, the difference between the Fe-Fe and Fe-$R$ disorder
parameters is significant when the cases of the light and heavy fillers
are compared. This concerns not only the temperature dependence of
the parameters, which is set by the relevant energy scales (either
the Debye or Einstein temperatures), but also the order of magnitude
of the parameters. In particular, one should note the relatively small
disorder of the Fe-Ba path in comparison with the Fe-K and Fe-Ca cases.
For a more quantitative analysis, we turn to the obtained energy scales.
From now on, we denote by $\theta_{D}^{R,\text{Fe-}Y}$ the Debye
temperature obtained for the Fe - $Y$ path of the $R$ filled sample
and by $\theta_{E}^{\text{Fe}-R}$ the analogous Einstein temperature.
All results are compiled in table \ref{tab:tableEXAFSparameters}.

The $\theta_{D}^{R,\text{Fe-Sb}_{1}}$ parameters display a weak dependence
with $R$ and compare well with those obtained in the early study
by Cao \emph{et al} \citep{cao_evidence_2004} for the $R=$ Ce and
La cases ( $\approx390$ K), suggesting that the Fe-Sb bondings are
not much affected by the filler. The $\theta_{E}^{\text{Fe}-R}$ here
obtained, however, are significantly larger than those obtained by
EXAFS of the $R=$ La and Ce cases, but are closer to values determined
from atomic displacement parameters (ADPs) for the $R=$ Ba and Ca
cases by \citet{schnelle_magnetic_2008} (one should, of course, keep
in mind the difference between the quantities obtained by EXAFS and
ADPs). As for the overall trend in the data, $\theta_{E}^{\text{Fe-Ba}}>\theta_{E}^{\text{Fe-K}}>\theta_{E}^{\text{Fe-Ca}}$,
similar results are obtained by heat capacity \citep{schnelle_magnetic_2008},
even though the heat capacity results are more indicative of $\theta_{E}^{\text{Fe-Ba}}>\theta_{E}^{\text{Fe-K}}\approx\theta_{E}^{\text{Fe-Ca}}$.
Concerning the $\theta_{D}^{R,\text{Fe-Fe}}$ parameters, the values
obtained in the $R=$ K or Ca cases are similar to results for other
skutterudites \citep{keiber_modeling_2015,bridges_complex_2015,cao_evidence_2004}
whereas the Fe-Fe vibrations are particularly harder in the $R=$
Ba case.

The high-temperature limit of the $\sigma^{2}$ parameters of each
single scattering path can be connected to an effective spring constant
($K_{\text{eff}}$) as discussed in detail in Refs. \citep{keiber_modeling_2015,bridges_complex_2015}.
We denote by $K_{\text{eff}}^{R,\text{Fe-}Y}$ the effective spring
constant for the Fe - $Y$ path of the $R$ filled sample. Effective
spring constants describe better the vibrational properties than a
direct comparison between the Debye (Einstein) temperatures since
they take into account the reduced mass of the vibrating elements
and the fact that the $K_{\text{eff}}^{R,\text{Fe-}Y}$ are not linearly
dependent on the Debye (Einstein) temperatures. The parameters are
listed in table \ref{tab:tableEXAFSparameters}.

\begin{table*}
\caption{The $\theta$ (either $\theta_{D}^{R,\text{Fe-}Y}$ or $\theta_{E}^{\text{Fe}-R}$
) and $K_{\text{eff}}$ parameters obtained from the analysis of the
correlated Debye-Waller parameters. \label{tab:tableEXAFSparameters}}

\centering{}%
\begin{tabular*}{1\textwidth}{@{\extracolsep{\fill}}ccccccc}
 & \multicolumn{2}{c}{} & \multicolumn{2}{c}{} & \multicolumn{2}{c}{}\tabularnewline
\hline 
\hline 
Fe $K$-edge & \multicolumn{2}{c}{KFe$_{4}$Sb$_{12}$} & \multicolumn{2}{c}{CaFe$_{4}$Sb$_{12}$} & \multicolumn{2}{c}{BaFe$_{4}$Sb$_{12}$}\tabularnewline
Path & $\theta$ (K) & $K_{\text{eff}}$ (eV/$\text{Å}^{2}$) & $\theta$ (K) & $K_{\text{eff}}$ (eV/$\text{Å}^{2}$) & $\theta$ (K) & $K_{\text{eff}}$ (eV/$\text{Å}^{2}$)\tabularnewline
\hline 
Fe - Sb$_{1}$ & 422(15) & 8.13 & 419(14) & 8.13 & 426(17) & 8.45\tabularnewline
Fe - \textit{R} & 168(61) & 1.15 & 134(51) & 0.74 & 190(43) & 2.55\tabularnewline
Fe - Sb$_{2,3}$ & 356(26) & 3.85 & 323(23) & 3.12 & 360(15) & 3.92\tabularnewline
Fe - Fe$_{1}$ & 395(75) & 3.56 & 390(77) & 3.45 & 620(80) & 9.79\tabularnewline
\hline 
\hline 
 & \multicolumn{2}{c}{} & \multicolumn{2}{c}{} & \multicolumn{2}{c}{}\tabularnewline
\end{tabular*}
\end{table*}

Translated to $K_{\text{eff}}^{R,\text{Fe-}Y}$, the vibrational properties
of Fe - Sb bondings for the $R=$ K, Ca and Ba filled materials are
similar, with the difference being less than $5\%$ between the parameters
$K_{\text{eff}}^{R,\text{Fe-Sb}_{1}}$. They are also very similar
to the vibrational properties of the Co - Sb bondings as investigated
by EXAFS \citep{rodrigues_atomic_2022}. Inspection of the $K_{\text{eff}}^{R,\text{Fe-\ensuremath{R}}}$parameters,
on the other hand, reveals a much larger variation as a function of
$R$. Indeed, $K_{\text{eff}}^{\text{Ba},\text{Fe-\text{Ba}}}/K_{\text{eff}}^{\text{Ca},\text{Fe-Ca}}\approx3.5$,
stating a clear distinction between the cases of heavy and light fillers.
It suggests a relative decoupling of the filler-cage vibrational dynamics
in the case of the light-weight cations, a result that is in agreement
with previous analysis of the $R$Fe$_{4}$Sb$_{12}$ atomic displacement
parameters (ADPs) \citep{schnelle_magnetic_2008}. It so appears that
the light-weight cations may behave as ``rattlers'' which, however,
are not totally independent of the cage vibrations. We note that the
$K_{\text{eff}}^{\text{Ba},\text{Fe-\text{Ba}}}$ parameter compares
well with the Ru - Ce and Pt - La cases, obtained from EXAFS experiments
of CeRu$_{4}$As$_{12}$ \citep{bridges_local_2016,keiber_modeling_2015}
and LaPt$_{4}$Ge$_{12}$ \citep{MCantarino_LaPt4Ge12_2022} respectively,
which are other examples of heavy fillers filled skutterudites. These
results, however, are not simply expressing the filler-cage mass relation,
but rather the potential energies of the cations, as discussed by
previous calculations of the $R$Fe$_{4}$Sb$_{12}$ \citep{koza_vibrational_2010}.
The Fe - Fe vibrations are evidenced by the analysis of the $K_{\text{eff}}^{R,\text{Fe-Fe}}$
parameters to be distinctively harder in the case of the Ba filled
samples, suggesting stronger Fe-Fe bonding in this case. We note that
$K_{\text{eff}}^{\text{Ba},\text{Fe-\text{Fe}}}/K_{\text{eff}}^{R,\text{Fe-Fe}}\approx$2.8
for both $R=$ K or Ca.

The effective spring constants can be applied to the discussion of
an effective $1$D phonon model, introduced by Keiber \emph{et al.}
\citep{keiber_modeling_2015,bridges_complex_2015}, for the qualitative
description of the correlated filler-cage motion in filled skutterudites.
The model considers the three distinct elements illustrated in figure
\ref{fig:structure_exafsresults}$(c)$: the $R$ filler (or rattler),
the Sb square ring and the Fe metal cage, which denotes solely the
Fe cubic lattice. The coupled character of the vibrational dynamics
is given by connecting the elements by four distinct spring constants
for each $R$ filled skutterudite: a rattler-square ring ($k_{\text{rs}}^{R}$),
a rattler-metal cage ($k_{\text{rc}}^{R}$), a metal cage-metal cage
($k_{\text{cc}}^{R}$) and a metal cage-square ring ($k_{\text{cs}}^{R}$).
The system dynamics is treated at a classical level and the corresponding
dynamic matrix is diagonalized to find four phonon dispersion modes
$\omega_{j}(q)$, where $q$ is the phonon wave-vector \citep{keiber_modeling_2015}.
We shall discuss two aspects of the coupled vibrations that can be
connected to our experimental results: $i)$ the filler-cage mass
relation; $ii)$ the dependence of he $k_{\text{rc}}^{R}$ constants
on whether $R$ is a light or heavy element (as suggested by the experimentally
obtained $K_{\text{eff}}^{R,\text{Fe-}R}$ parameters). Moreover,
we shall fix the parameter $k_{\text{cs}}^{R}=$1.8 eV/$\text{Å}^{2}$
for all samples, as suggested by the experimentally obtained $K_{\text{eff}}^{R,\text{Fe-Sb}}$
and adopt$k_{\text{cc}}^{\text{K(Ca)}}=10$ eV/$\text{Å}^{2}$and
$k_{\text{cc}}^{\text{Ba}}=25$ eV/$\text{Å}^{2}$based on the variation
of the obtained $K_{\text{eff}}^{R,\text{Fe-Fe}}$ parameters. Our
experiments do not offer an estimate about the $k_{\text{rs}}^{R}$
parameters and we chose to adopt a range of values in reference to
previous works \citep{keiber_modeling_2015,bridges_local_2016}.

\begin{figure}
\begin{centering}
\includegraphics[width=1\columnwidth]{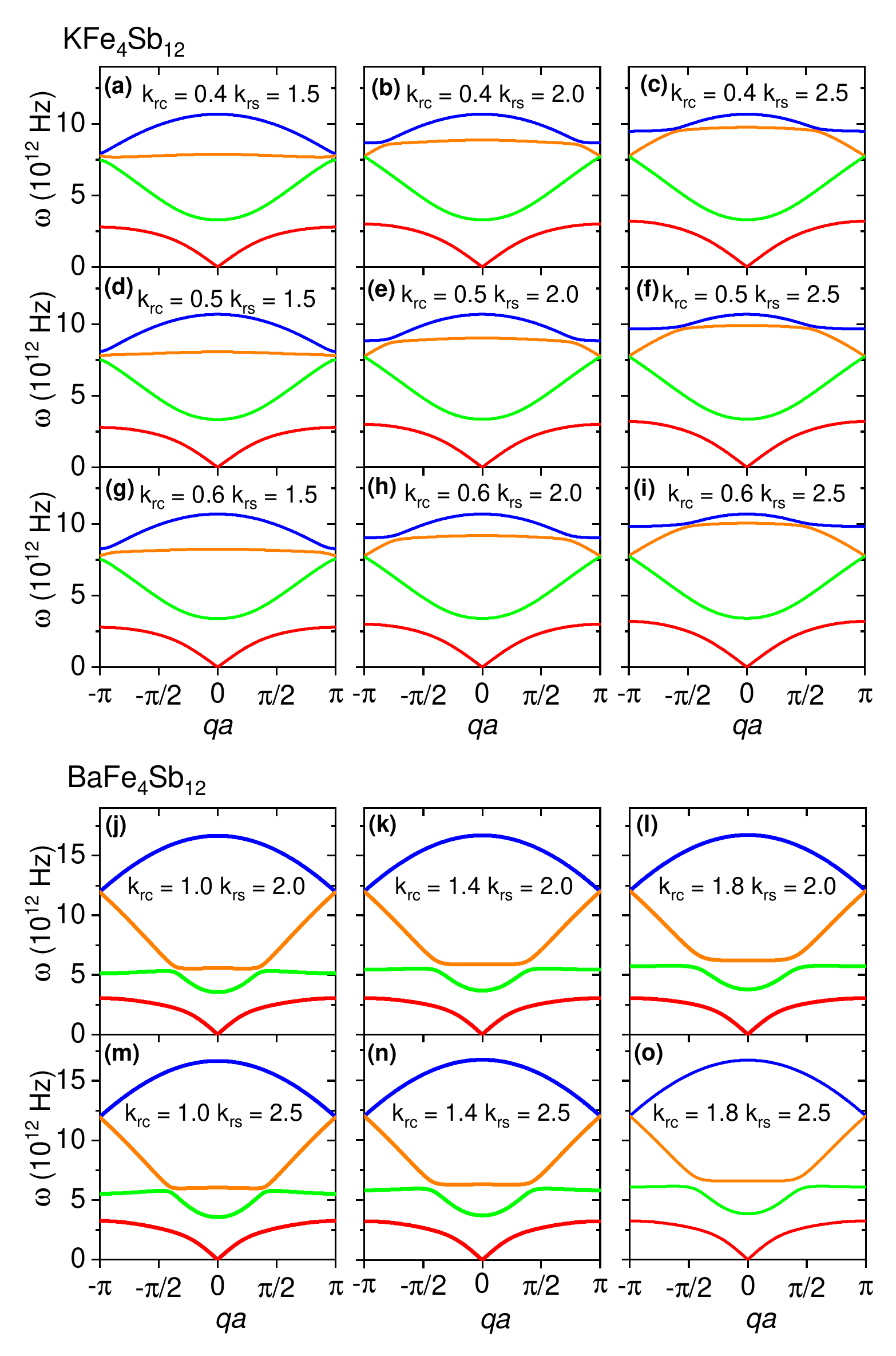}
\par\end{centering}
\caption{Effective 1D correlated rattling model for the $R=$ K (Ca) and Ba
samples. In ascending frequency, we have the acoustic mode (red) and
the first, (green), second (orange) and third (blue) optical phonon
modes. The modes are displayed as function of $qa$ where $a$ is
the material lattice constant and $q$ is the phonon wave-vector.
In panels $(a)-(i)$, the mass relation between the vibrating elements
is that of the $R=$ K case and in panels $(j)-(o)$, the mass relation
between the vibrating elements is that of the $R=$ Ba case. The adopted
parameters are: $k_{\text{cs}}^{\text{R}}=1.8$ eV/$\text{Å}^{2}$
for all panels and $k_{\text{cc}}^{\text{K}}=10$ eV/$\text{Å}^{2}$,
in panels $(a)$-$(i)$, and $k_{\text{cc}}^{\text{Ba}}=25$ eV/$\text{Å}^{2}$,
in panels $(j)$-$(o)$, reflecting the different $K_{\text{eff}}^{R,\text{Fe-Fe}}$
parameters. The adopted $k_{\text{rs}}^{R}$ and $k_{\text{rc}}^{R}$
values are indicated in each panel in units of eV/$\text{Å}^{2}$.
\label{fig:correlatedDebye}}
\end{figure}

One should keep in mind that there is no one-to-one correspondence
between the $K_{\text{eff}}^{R,\text{Fe-}Y}$ and spring constants
of the 1D effective model. Rather, the $K_{\text{eff}}^{R,\text{Fe-}Y}$
constants are adopted to get reasonable estimates to the spring constants,
in particular to their relative values. Formally, the spring constants
are a functional of the $K_{\text{eff}}^{R,\text{Fe-}Y}$ and there
is a strong functional dependence between, for instance, $K_{\text{eff}}^{R,\text{Fe-Fe}}$
and $k_{\text{cc}}^{R}$ or $K_{\text{eff}}^{R,\text{Fe-}R}$ and
$k_{\text{rc}}^{R}$. Results of the phonon dispersion and the exact
parameters adopted are displayed in figures \ref{fig:correlatedDebye}$(a)$-$(o)$
and in the figure caption.

In figures \ref{fig:correlatedDebye}$(a)$-$(i)$, the mass-relation
between the $R$ cations and the cage elements is that of the KFe$_{4}$Sb$_{12}$
case (which, in view of the qualitative nature of the calculated dispersions,
also describe the CaFe$_{4}$Sb$_{12}$ case) and is one of the key
properties determining the phonon dispersion. The other relevant parameter
is indeed the relatively small rattler-cage spring constant. These
two parameters determine, respectively, the wide separation between
the acoustic and optical modes and the conspicuous flattening of the
second optical mode (thick orange line). The latter implies a wide
range of $q$-values for which the phonon group velocity is very close
to zero, describing a rather localized phonon mode, reminiscent of
the Einstein oscillators scenario. In view of the ratio $K_{\text{eff}}^{\text{Ca},\text{Fe-Ca}}/K_{\text{eff}}^{\text{K},\text{Fe-K}}\approx2/3$
(see table \ref{tab:tableEXAFSparameters}) we propose that the CaFe$_{4}$Sb$_{12}$
and KFe$_{4}$Sb$_{12}$ vibrational dynamics are, respectively, described
by figures \ref{fig:correlatedDebye}$(a)$-$(c)$ and \ref{fig:correlatedDebye}$(g)$-$(i)$.
The range of values in the $k_{\text{rs}}^{R}$ leaves the acoustic
and first optical modes unchanged (red and green lines), while pushing
the second optical mode (orange) up, causing this mode to interact
with the third one (blue).

The mass relation of the BaFe$_{4}$Sb$_{12}$ case is represented
in figures \ref{fig:correlatedDebye}$(j)$-$(o)$. We now observe
a smaller frequency gap between the acoustic and optical modes (due
to the filler-cage mass relation) and a much smaller region wherein
the second optical mode is flat, due to the large rattler-cage coupling.
Here, the parameters' range is similar to the one adopted in the CeRu$_{4}$As$_{12}$
\citep{keiber_modeling_2015} and LaPt$_{4}$Ge$_{12}$ \citep{MCantarino_LaPt4Ge12_2022}
cases, which allows us to pinpoint that the filler-cage mass relation
is also responsible for a larger flattening of the optical mode, contributing
to the phonon localization as recently proposed \citep{valerio_phonon_2020}.
The value of the $k_{\text{cc}}^{\text{Ba}}$ constant (reflecting
the large $K_{\text{eff}}^{\text{Ba},\text{Fe-\text{Fe}}}$) is twice
the value for the other heavy-filler filled skutterudites and it essentially
increases the frequency of the higher energy (blue) optical mode.
In comparison to the K (Ca) case, this shift of the third optical
mode to a higher frequency causes the decoupling of this mode from
the second optical mode, which in the Ba case turns out to interact
mainly with the first optical mode.

\subsection{X-ray diffraction and elastic properties}

\begin{figure*}
\begin{centering}
\includegraphics[width=1\textwidth]{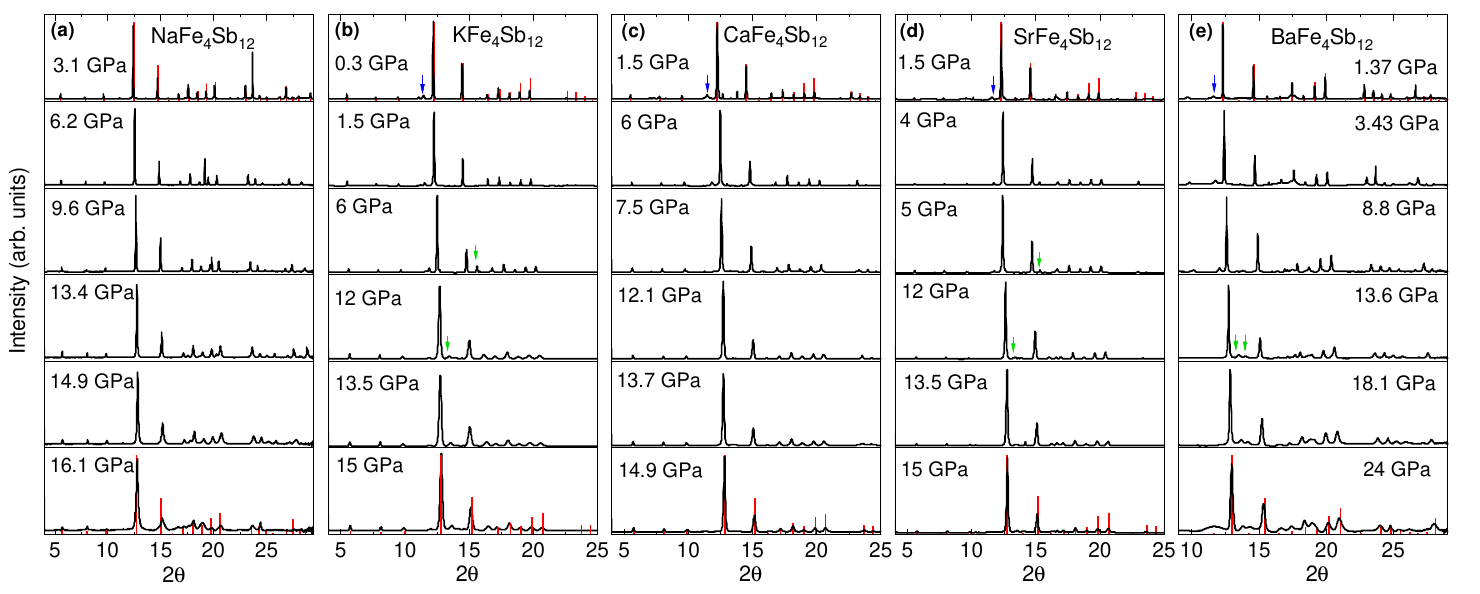}
\par\end{centering}
\caption{Representative $P$ dependent XRD experiments for all $R$Fe$_{4}$Sb$_{12}$
filled skutterudite, wherein we have: $(a)$ $R=$ Na; $(b)$ $R=$
K; $(c)$ $R=$ Ca; $(d)$ $R=$ Sr; and $(e)$ $R=$ Ba. The $P$
is as indicated in the panels. The thick red lines represent the diffraction
peaks of the skutterudite structure. The blue and green arrows point
to diffraction peaks not ascribed to the skutterudite phase and are
discussed in the main text and in the supplemental material. \label{fig:XRD_Results_experiments}}
\end{figure*}

We now turn to the $P$ dependent XRD experiments. In figures \ref{fig:XRD_Results_experiments}$(a)-(e)$
representative data of the XRD experiments, after removing the background,
are presented for all samples. The main diffraction peaks of the skutterudite
structure are marked in red as a reference. The powder profile was
compared to the skutterudite crystal structure (the thick red lines
in figures \ref{fig:XRD_Results_experiments}$(a)-(e)$ ) and the
$P$ dependence of the lattice parameters was extracted from the peak
positions. The blue and green arrows point to peaks that are not part
of the skutterudite phase. The peaks are, respectively, ascribed to
small amounts of non-reacted Sb and to the sample decomposition at
high \emph{P} (see supplemental material).

\begin{figure}
\begin{centering}
\includegraphics[width=1\columnwidth]{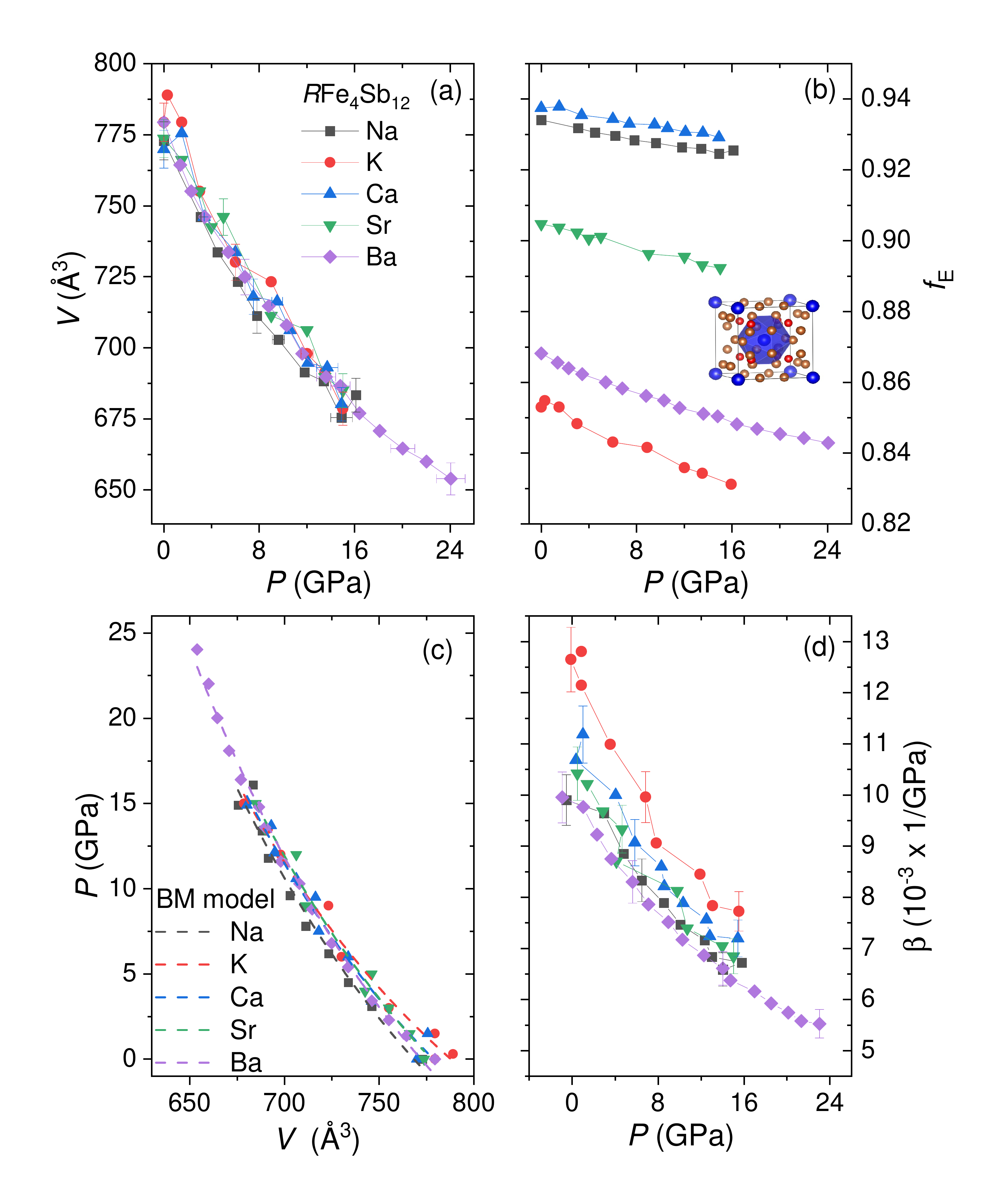}
\par\end{centering}
\caption{$(a)$ The unit cell volume $V$ as function of $P$. Error bars for
$P$ are estimated from the half-width at half maximum of the ruby
fluorescence spectra and amount to about $3-5$ $\%$ in this pressure
interval. $(b)$ The cage empty volume fraction $f_{\text{E}}$ (see
text) as a function of $P$. The inset shows a representation of the
skutterudite structure highlighting the Sb cage around the filler
site (the cage volume was calculated with VESTA \citep{momma_vesta_2008}).
$(c)$ The $P$ as a function of $V$ data and their respective fittings
to the Birch-Murnaghan model. $(d)$ The compressibility $\beta$
as a function of $P$ for all samples. \label{fig:XRD_results_parameters}}
\end{figure}

Our analysis of the XRD experiments are in figures \ref{fig:XRD_results_parameters}$(a)-(d)$.
In figure \ref{fig:XRD_results_parameters}$(a)$, the unit cell volumes
($V$) as a function of $P$ of all samples are presented. It is tempting
to associate the skutterudite elastic properties to the relation between
the cage volume and the filler cationic size. We then compile in figure
\ref{fig:XRD_results_parameters}$(b)$ a phenomenological parameter:
the cage empty volume fraction $f_{E}$, that we define as $f_{E}=1-V_{\text{R}}/V_{\text{C}}$,
where $V_{\text{R}}$ is the $R$ cationic volume and $V_{\text{C }}$is
the Sb cage volume, as a function of $P$. The cage is represented
in the inset of figure \ref{fig:XRD_results_parameters}$(b)$ as
a reference.

The volume pressure dependence is investigated by fitting $V\times P$
data to the Birch-Murnaghan (BM) model \citep{birch_finite_1947}:

\begin{equation}
P(V)=\frac{3}{2}B_{0}[(\frac{V}{V_{0}})^{-\frac{7}{3}}-(\frac{V}{V_{0}})^{-\frac{5}{3}}]\{1+\frac{3}{4}(B_{0}'-4)[(\frac{V}{V_{0}})^{-\frac{2}{3}}-1]\}\label{eq:birchEquation}
\end{equation}

where $V$ is the sample unit cell volume, $V_{0}$ is the sample
volume at room conditions, $B_{0}$ is the bulk modulus and $B_{0}'$
is the bulk modulus derivative. We keep $B_{0}'=4$ GPa, as recently
adopted to describe other skutterudites \citep{sergueev_quenching_2015,rodrigues_unveiling_2021}
in this pressure range. Values obtained for $V_{0}$ and $B_{0}$
are compiled in table \ref{tab:tablePressureParameters}. The $V_{0}$
values obtained from the fitting deviate only about $1$\% from the
experimentally determined from x-ray diffraction at room conditions
suggesting the adequacy of our analysis. The $V\times P$ data and
their respective fittings are in figure \ref{fig:XRD_results_parameters}$(c)$.

The $B_{0}$ values thus obtained range from $77$ GPa ($R=$ K) to
$99$ GPa ($R=$ Ba). The values compare well with those obtained
for other $R$Fe$_{4}$Sb$_{12}$ based skutterudites, such as EuFe$_{4}$Sb$_{12}$
\citep{sergueev_quenching_2015} and CeFe$_{4}$Sb$_{12}$ \citep{shirotani_x-ray_2004,liu_synchrotron_2011}.
The latter further suggests that $B_{0}$ reflects the rigidity of
the Fe$_{4}$Sb$_{12}$ framework which, however, is not in total
$R$ independent. The experimentally obtained $B_{0}$ parameters
are in the range of the values previously obtained by calculations
of the $R=$ Ca, Sr and Ba cases \citep{koza_vibrational_2011}. It
is noteworthy, however, that the calculated values are nearly $R$
independent and about $80$ GPa. Our results also compare well with
other Sb-based skutterudites, such as CoSb$_{3}$ \citep{rodrigues_unveiling_2021},
in agreement with the understanding that the pnicogen square ring
rigidity dominates the skutterudite elastic properties. Indeed, the
wider variation in $B_{0}$ is observed when the pnicogen is changed,
with P forming characteristically rigid cages \citep{shirotani_x-ray_2004}.

\begin{table}
\caption{Parameters of the Birch-Murnaghan model for all $R$Fe$_{4}$Sb$_{12}$
samples. \label{tab:tablePressureParameters}}

\centering{}%
\begin{tabular*}{1\columnwidth}{@{\extracolsep{\fill}}ccccccccccc}
 &  &  &  &  &  &  &  &  &  & \tabularnewline
\hline 
\hline 
$R=$ & \multicolumn{2}{c}{Na} & \multicolumn{2}{c}{K} & \multicolumn{2}{c}{Ca} & \multicolumn{2}{c}{Sr} & \multicolumn{2}{c}{Ba}\tabularnewline
\hline 
$V_{0}$($\textrm{Å}{}^{3}$) & \multicolumn{2}{c}{768.3(7)} & \multicolumn{2}{c}{787.8(2)} & \multicolumn{2}{c}{778.7(5)} & \multicolumn{2}{c}{777.6(5)} & \multicolumn{2}{c}{772.0(3)}\tabularnewline
$B_{0}$(GPa) & \multicolumn{2}{c}{94(7)} & \multicolumn{2}{c}{77(5)} & \multicolumn{2}{c}{87(6)} & \multicolumn{2}{c}{91(6)} & \multicolumn{2}{c}{99(3)}\tabularnewline
\hline 
\hline 
 & \multicolumn{2}{c}{} & \multicolumn{2}{c}{} & \multicolumn{2}{c}{} & \multicolumn{2}{c}{} & \multicolumn{2}{c}{}\tabularnewline
\end{tabular*}
\end{table}

In possession of the $B_{0}$ values, we then obtain the materials'
compressibility $\beta$ as a function of $P$. The results are displayed
in figure \ref{fig:XRD_results_parameters}$(d)$. The compressibility
data express well how the elastic properties depend on $R$: it is
shown that the K and Ca filled materials have larger $\beta$ while
the Ba and Na filled materials have smaller $\beta$. It is a natural
assumption to speculate that the higher $f_{\text{E }}$ more compressible
would be the material. If the whole data is inspected, this trend
is not obeyed. For instance, the Ca filled material, which has the
largest $f_{E}$, is not the most compressible. If, however, we focus
on either the $R^{1+}$ or $R^{2+}$ filled materials, an interesting
picture emerges. Restricting the analysis to the case of the $R^{1+}$
fillers, one notes that this simple picture does not work and $\beta$
does not correlate with $f_{E}$, with the K filled material being
the more compressible. Now, if one inspects the case $R^{2+}$ fillers
case, the experimentally obtained picture obeys the proposed trend,
with the Ca filled material being the more compressible.

The case of the alkaline earth filled materials is distinct due to
the presence of the heavy Sr and Ba fillers. Here, Sr and Ba virtual
atomic $4d$ and $5d$ states, respectively, are more extended in
real space as well as energetically accessible than in the case of
the Ca virtual $3d$ states. Thus, the $4d$ and $5d$ states may
take part in the filler-cage bonding which, in this scenario, is not
completely ionic in character, as has been proposed for La filled
skutterudites \citep{grosvenor_x-ray_2006}. To verify this hypothesis,
we performed DFT calculations of the filler-cage bonding properties.

We focus our analysis on the DFT results on the filler derived atomic
orbitals (AOs) contribution to the molecular orbitals (MOs) with the
highest superposition with the Sb-derived $5p$ states, since these
are precisely the MOs contributing the most to the filler--cage bonding.
These MOs sit in the vicinity of the Fermi Level (see supplemental material).
Our results show that whereas the $4d$ and $5d$ virtual AOs from
Sr and Ba, respectively, take part in the formation of these MOs,
Ca only contributes its $3p$ states to the bonding MOs. In table
\ref{tab:orbitalbonding}, we list the filler derived AOs contributing
the most (largest superposing orbitals) to the formation of the filler-cage
bonding MOs. The MOs symmetry classification is indicated and is made
in terms of the irreducible representations of the $T_{h}$ group,
which is the point group symmetry of the $R$ site \citep{takegahara_crystal_2001,garcia_direct_2008}.
The results were obtained at room conditions and at high $P$ ($\approx13.6$
GPa). The $P$ effect was simulated by adopting the results from our
XRD experiments.

\begin{table*}
\caption{The highest filler derived superposing AOs to the formation of the
MOs connected to the filler-cage bonding. The nature of the AOs and
the symmetry character of the MOs are indicated. The Final orbital
composition analysis was performed adopting the Ros-Schuit partition
method via the Multifwn program \citep{lu_multiwfn_2012}. \label{tab:orbitalbonding}}

\centering{}%
\begin{tabular}{cccccc}
 &  &  &  &  & \tabularnewline
\hline 
\hline 
\multirow{2}{*}{$R$} & \multirow{2}{*}{AOs character} & \multicolumn{2}{c}{Room Pressure} & \multicolumn{2}{c}{$\mathrm{\sim13.6}$ GPa}\tabularnewline
\cline{3-6} \cline{4-6} \cline{5-6} \cline{6-6} 
 &  & Orbital Cont. (\%) & $\mathrm{E-E_{F}}$ (eV) & Orbital Cont. (\%) & $\mathrm{E-E_{F}}$ (eV)\tabularnewline
\hline 
Ca & $3p_{x}$ $(t_{u})$ & 30.77 & $-3.23$ & 30.82 & $-3.46$\tabularnewline
Sr & $4d_{xy},4d_{xz},4d_{yz}$ $(t_{g})$ & 11.84 & $-1.70$ & 15.37 & $-1.82$\tabularnewline
Ba & $5d_{xy},5d_{xz},5d_{yz}$ $(t_{g})$ & 19.54 & $-1.60$ & 28.00 & $-1.68$\tabularnewline
\hline 
\hline 
 &  &  &  &  & \tabularnewline
\end{tabular}
\end{table*}

As can be observed, pressure has nearly no effect on the bonding in
the case of the Ca filler, whereas it further stabilizes and increases
the $d$ states participation in the Sr and Ba filled materials. This
effect is particularly large in the case of Ba, for which an increase
of more than $\gtrsim40\%$ of the $5d$ orbital contribution to MOs
enrolled in the filler-cage bonding is observed in the high $P$ calculations.
This result means that on top of geometrical considerations, one must
consider that $P$ induces further stabilization to the filler-cage
bonding via the available virtual $nd$ states, rendering the material
more rigid. Our findings thus suggest that the association between
$B_{0}$ (or $\beta$) and $f_{\text{E}}$ in the case of the $R^{2+}$
fillers actually reflects a deeper understanding of the problem based
upon the filler-cage bonding properties, in particular in the case
of the heavy fillers. This mechanism could also be at work in the
context of the partially filled $M_{y}$Co$_{4}$Sb$_{12}$ skutterudites,
for which it is observed that when $M$ is a heavy filler, in particular
in the cases of $M=$ La and Yb, the material is more rigid \citep{rodrigues_unveiling_2021}.
Moreover, our calculations suggest an association between the participation
of the Ba $5d$ states in the filler-cage bonding and the relatively
large value of the $K_{\text{eff}}^{\text{Fe-Ba}}$ parameter.

\begin{figure*}
\includegraphics[width=1\textwidth]{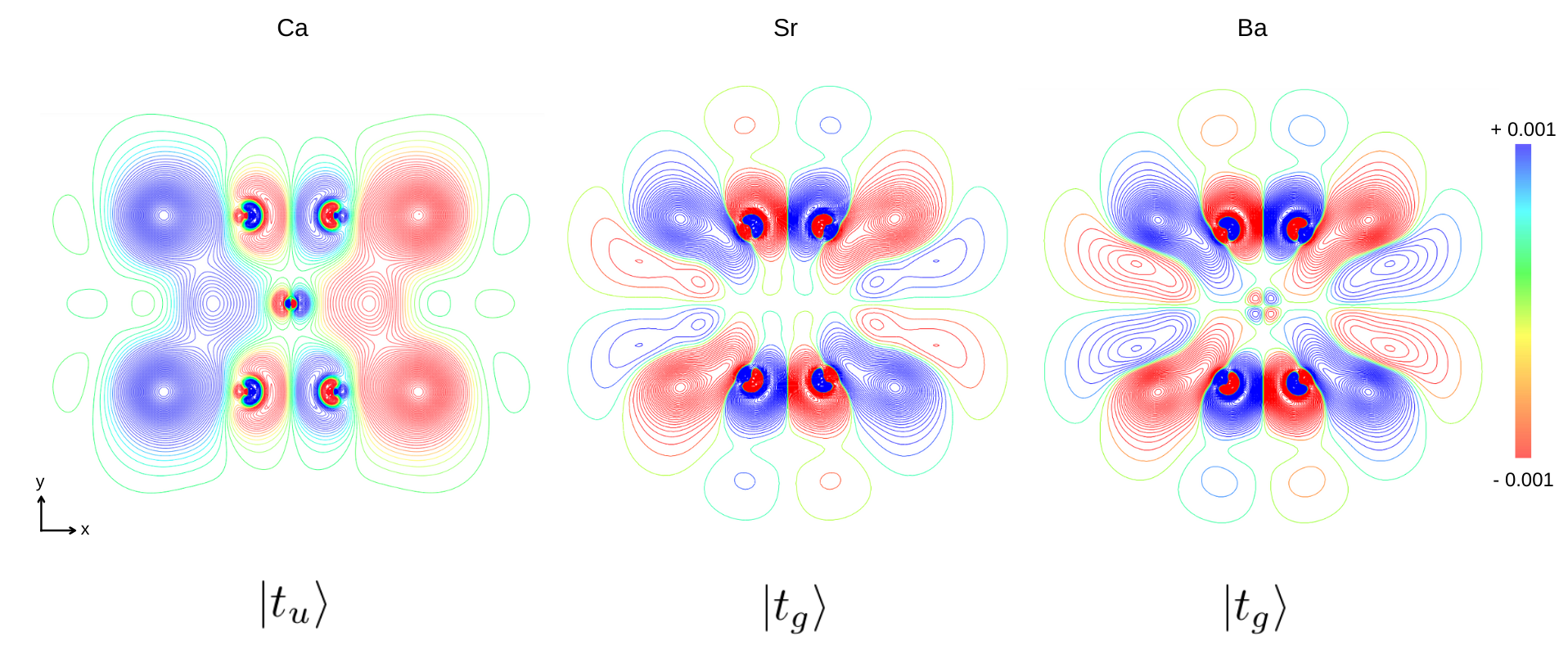}

\caption{Contour plots for the MOs describing the filler-cage bonding in the
$(x,y)$ plane in the cases of the Ca, Sr and Ba filled materials
as indicated. The figures are formed by $120$ contours in the isovalue
interval indicated in the color scale in the right side of the figure.
Ligand Orbitals are plotted via the Gabedit program \citep{allouche_gabeditgraphical_2011}.
\label{fig:bondingfigure}}
\end{figure*}

A better visualization of the filler-cage bonding, in particular of
the $d$-character of the MOs in the Sr and Ba case, can be observed
in figure \ref{fig:bondingfigure}, wherein we depict the MOs related
to the results listed in table \ref{tab:orbitalbonding}. The plots
are contour plots of the mentioned MOs and illustrate the real space
electron density distribution around the filler atom, which is located
at the center of the figures. As it is clear, the plotted MOs are
dominated by orbitals of $d$-character in the case of Sr- and Ba-filled
materials, whereas in the case of Ca-filled material the presence
of the $3p$ lobe dominating the MOs is clearly distinguishable.

Lastly, we comment on the significant broadening of the diffraction
peaks for high pressures that is observed. Starting at about $12$
GPa in the $R=$ Na and K cases and at higher $P$ in the case of
the alkaline earth filled skutterudites, the peaks' broadening amount
to an increasing structural disorder induced by $P$. The alkaline
earth filled materials are suggested to be relatively less susceptible
to the disorder increase by $P$. Indeed, the diffraction peaks can
still be observed for $P$ up to $24$ GPa in the BaFe$_{4}$Sb$_{12}$
case. The skutterudite structure is mainly stabilized by a combination
of geometric and charge constraints related to the filler and cage
sizes and charge distribution \citep{gumeniuk_filled_2010,luo_large_2015}.
We thus conclude that the larger structural stability of alkaline
earth filled materials is likely due to the bonding stabilization
of the extra charge offered by the $R^{2+}$ cations.

\section{Summary and Conclusions}

The site specific vibrational properties of the $R=$ K-, Ca- and
Ba-filled skutterudites $R$Fe$_{4}$Sb$_{12}$ were investigated
by EXAFS experiments. As concluded from the effective spring constants
obtained from experiments, materials filled with light-weight cations
display weak rattler-cage couplings, reminiscent of the independent
rattler scenario, and the $R=$ Ba material displays a characteristic
strong Fe-Fe coupling. Based upon an effective $1$D model for the
skutterudite phonon dispersion, we proposed that the $R=$ K- and
Ca-filled skutterudites should display a conspicuous optical flat
mode in accordance with the Einstein oscillator scenario.

We thus introduced our pressured dependent experiments. The $V\times P$
curves were determined and fitted to the Birch-Murnaghan model to
extract the bulk modulus $B_{0}$ and then the compressibility $\beta$.
We found that the geometric parameter $f_{\text{E}}$ cannot explain
the $B_{0}$ (or $\beta$) dependence as a function of $R$. A geometrical
relation, however, was observed to hold in the case of the $R^{2+}$
filled materials. By considering DFT calculations, we uncovered that
this association between $B_{0}$ (or $\beta$) and $f_{\text{E}}$
actually reflects an emerging property of the filler-cage bonding,
which is particularly relevant for heavy fillers. Moreover, the filler-cage
bonding can also explain the large $K_{\text{eff}}^{\text{Fe-Ba}}$
parameter, since the contribution from Ba $5d$ states to bonding
makes the Ba more tightly coupled to the cage (no ``independent''
rattler in this case).

From the point of view of material design, our results suggest that
synthesizing mixed filled skutterudites, featuring light and heavy
fillers, is a good strategy to introduce localized vibrational modes
(``rattlers'') in the material's vibrational dynamics. This speculation
has some ground on our calculations showing that the bonding scheme
in skutterudites may include a certain degree of covalency in the
case of heavy fillers. Thus, introducing light cations into materials
mainly filled with heavy cations create a situation wherein the light
fillers are weakly bonded and sitting in an oversized cage.
\begin{acknowledgments}
The authors acknowledge CNPEM-LNLS for the concession of beam time
(proposals No. $20160180$, No. $20160181$, No. $20170709$ and No.
$20190018$). The XAFS2 and XDS beamlines staff are acknowledged for
the assistance during the experiments. The Fundação de Amparo à Pesquisa
do Estado de São Paulo financial support is acknowledged by M.R.C.
(grant No. 2019/05150-7 and grant No. 2020/13701-0) and F.A.G. (grant
No. 2019/25665-1). This study was financed in part by CAPES - Finance
Code 001. E.M.B. acknowledges Fundação Carlos Chagas Filho de Amparo
à Pesquisa do Estado do Rio de Janeiro (grant No. E-26/202.798/2019).
A.L.J. thanks J. Grin for his steady support and interest in this
work.
\end{acknowledgments}

\bibliographystyle{apsrev4-2}
\bibliography{2021JuAbrantes_vibrationalElastic_references}

\end{document}

\title{Supplemental material for: Vibrational and structural properties of
the $R$Fe$_{4}$Sb$_{12}$ ($R=$Na, K, Ca, Sr, Ba) filled skutterudites}
\author{Juliana G. de Abrantes$^{1}$, Marli R. Cantarino$^{1}$, Wagner R.
da Silva Neto$^{1,2}$, Victória V. Freire$^{1}$, Alvaro G. Figueiredo$^{1}$,
Tarsis M. Germano$^{1}$, Bassim Mounssef Jr.$^{2}$, Eduardo M. Bittar$^{3}$,
Andreas Leithe-Jasper$^{4}$, Fernando A. Garcia$^{1}$}
\affiliation{$^{1}$Intituto de Física, Universidade de São Paulo, 05508-090, São
Paulo-SP, Brazil}
\affiliation{$^{2}$Instituto de Química, Universidade de São Paulo, 05508-090,
São Paulo-SP, Brazil}
\affiliation{$^{3}$Centro Brasileiro de Pesquisas Físicas, Rio de Janeiro, RJ
22290-180, Brazil}
\affiliation{$^{4}$Max Planck Institute for Chemical Physics of Solids, D-01187
Dresden, Germany.}
\begin{abstract}
Here we present the powder X-ray diffraction data for all $R$Fe$_{4}$Sb$_{12}$
($R=$ Na, K, Ca, Sr, Ba) and details of the quantum chemistry calculations
of the filler-cage bonding properties in the $R$Fe$_{4}$Sb$_{12}$
($R=$ Ca, Sr, Ba) filled. 
\end{abstract}
\maketitle

\section{X-ray diffraction}

\begin{figure}
\begin{centering}
\includegraphics[scale=0.5]{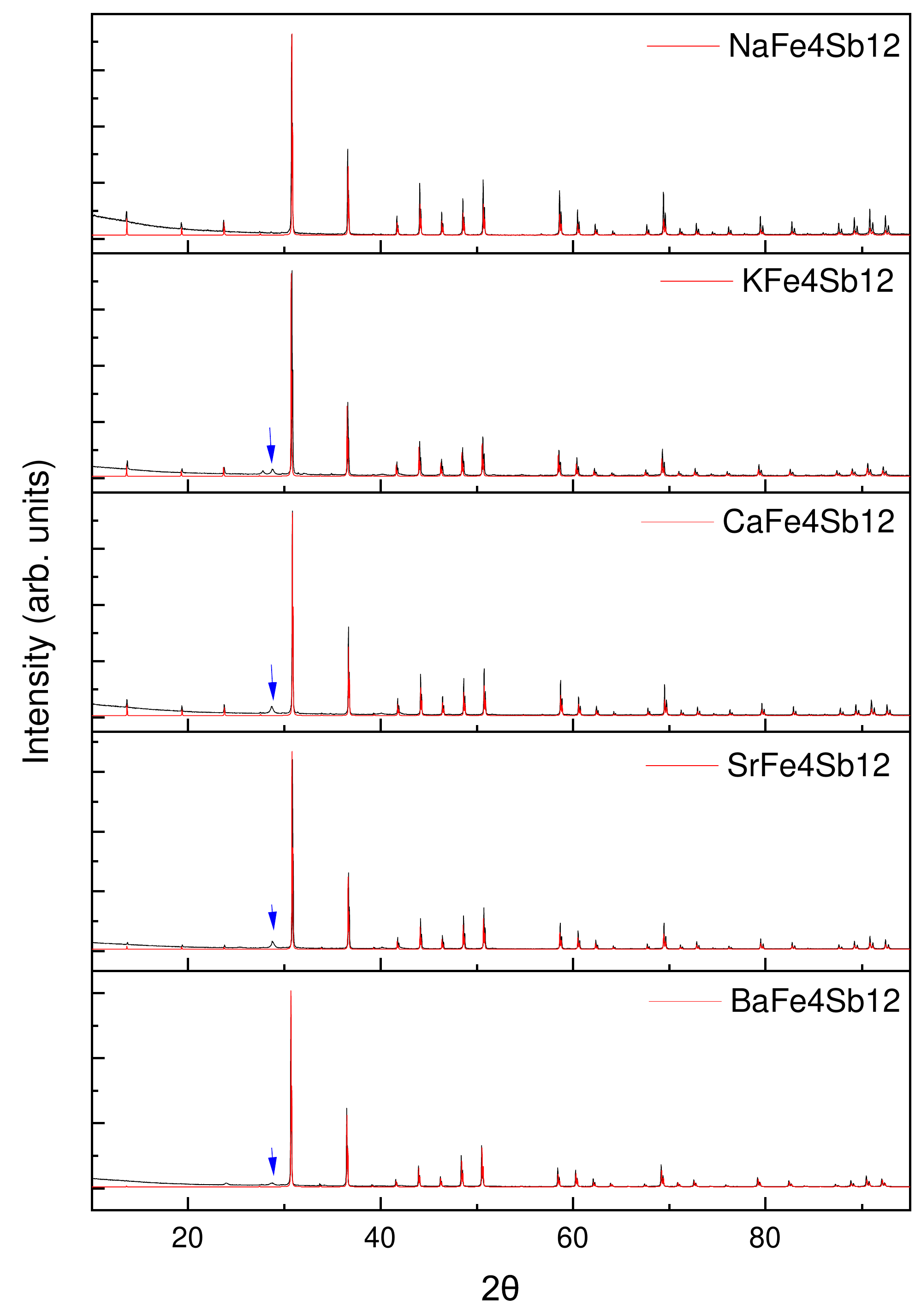}
\par\end{centering}
\caption{$R$Fe$_{4}$Sb$_{12}$ ($R=$ Na, K, Ca, Sr, Ba) at room conditions.
The lines in red show the skutterudite phase profile, while in black
is the collected data. The arrows in blue point the peaks in the diffraction
pattern attributed to an excess of non-reacted antimony, descendant
from the synthesis process. \label{fig1}}
\end{figure}

In Figure \ref{fig1} we show the X-ray diffraction pattern for all
the $R$Fe$_{4}$Sb$_{12}$ ($R=$ Na, K, Ca, Sr, Ba) filled skutterudites.
The materials were synthesized as described in Refs. \citep{leithe-jasper_weak_2004,leithe-jasper_ferromagnetic_2003}
by solid state reaction. The experiments were performed at room conditions,
employing Cu$K_{\alpha}$ radiation from a conventional X-ray tube.
All data were collected with a Bruker D8 diffractometer. The data
(thick black lines) are compared with the profile for the pure skutterudite
phase (thick black lines) and the results are compatible with high-quality
polycrystalline samples, with a small contribution of non-reacted
Sb or a minute contribution from non-skutterudite phase. 

\section{DFT calculations}

As discussed in the main text, in the case of the alkaline earth filled
materials ($R=$ Ca, Sr or Ba), the trend of the materials' bulk modulus
$B_{0}$ (or of their calculated compressibility $\beta$) cannot
be associated to a simple geometric parameter $f_{\text{E}}$. We
thus resorted to quantum chemistry calculations of the filler-cage
bonding properties to understand the evolution of $B_{0}$ as function
of $R$ in the series Ca $\rightarrow$ Sr $\rightarrow$ Ba. The
calculations were performed within the density functional theory approximation.
The building blocks of our quantum chemistry calculations are the
cage structure, formed by the $\mathrm{FeSb_{6}}$ octahedral complexes
and the $\mathrm{Sb_{4}}$ rings, plus the filler $R$. The electronic
states related to the filler-cage bonding were classified according
to the irreducible representations of the filler symmetry point group
$T_{h}$ \citep{takegahara_crystal_2001,garcia_direct_2008}. 

Within this description, we performed quantum chemistry calculations
of the filler-cage bonding properties using the BP86 functional for
all atoms. The ORCA 5.0 program package \citep{Neese2020,Neese2012a}
was adopted in the calculations. To better describe the local bonding
in the crystal structure, an \emph{embedded approach} was implemented
\citep{Dittmer2019}. A supercell with $339$ atoms was generated
among which $44$ cage atoms ($\mathrm{Fe_{8}Sb_{36}}$) plus one
$R^{2+}$ filler were treated in a quantum level. These $45$ atoms
were capped by $2$ atomic layers forming an effective core potential
(cECPs) to avoid spurious electron leakage. The remaining atoms were
simulated in a molecular mechanics fashion, with their charges ascribed
in agreement with the Zintl phase concept, constraining the number
of electrons as $24$ per $\mathrm{FeSb_{3}}$ unit \citep{gumeniuk_filled_2010,Luo2015},
by means of a force field. All atomic positions were obtained from
crystallographic data. For all atoms, the Karlsruhe valence triple-$\zeta$
with one set of polarization functions were adopted as the basis set,
as well as its auxiliary basis set for $J$ and $JK$ to speed up
calculations with the help of the $AutoAux$ function. Spin-orbit
coupling effects were considered by using the all-electron basis set
$\mathrm{x2c-TZVP}$ but no sizable effects were observed in the constitution
of the atomic orbitals (AOs) in the most prominent bonding molecular
orbitals (MOs).

Our results show that the $\left(n+1\right)d$ AOs ($n$ being the
highest occupied principal quantum number in Sr and Ba) are employed
as an additional stabilizing mechanism in the $\mathrm{Fe_{8}Sb_{36}}-R^{2+}$
bonding (and also in $\mathrm{Fe-Sb}$ bonding), defining $\mathrm{Fe_{8}Sb_{36}}-R^{2+}$
($R=$ Sr or Ba) as honorary complexes. The splitting of the MOs related
to the $\mathrm{Fe_{8}Sb_{36}}-R^{2+}$ bonding is larger than the
MOs splitting for the $3d-5p$ interactions. This large splitting
pushes the bonding MOs of the honorary complexes to a region in the
vicinity (and below) $E_{F}$, explaining why the $4d$ and $5d$
AOs from Sr and Ba, respectively, do not contribute to the Fermi surface. 

As there are several combinations for all MOs involved in the $\mathrm{Fe_{8}Sb_{36}}-R^{2+}$
bonding, a better inspection is made by plotting the orbital and element
specific contribution to the density of states (DOS). This is performed
via the Multifwn program \citep{lu_multiwfn_2012}. To visualize the
participation of the $d$ wavefunctions and Sb $5p$ orbitals in the
probed MOs, the Overlap population Density of States (OPDOS) was plotted
in figure \ref{fig:Total-DOS,-partial} (green lines), with the OPDOS
being defined as in eq. \ref{eq:OPDOS}:

\begin{equation}
\text{OPDOS}=\sum_{i}\left(c_{a,i}^{2}+\frac{c_{a,i}^{2}}{\sum_{b}c_{b,i}^{2}}\sum_{a}\sum_{b\neq a}c_{a,i}c_{b,i}\left\langle S_{ab}^{i}\right\rangle \right)\delta(E-\epsilon_{i})\label{eq:OPDOS}
\end{equation}
where $c_{a,i}$ are the linear combination coefficients for an atomic
orbital $a$ involved in the $i$-th molecular orbital, $S_{ab}^{i}$
is the overlap integral between AOs $a$ and $b$ for each $i$-th
molecular orbital with energy $\epsilon_{i}$. OPDOS thus gives the
normalized amount of superposition (overlap) of AOs $a$ and $b$
in the $i$-th Molecular orbital, expressing the weighted participation
of the specific AOs $a$ and $b$ in the stabilization (via the superposition
principle) of the $i$-th MO. In our plots, the delta functions were
described as a Gaussian function with a full width at half maximum
(FWHM) of about $1.3$ eV. 

In figure \ref{fig:Total-DOS,-partial}, we show the Total DOS and
the OPDOS due to Fe $3d$ and Sb $5p$ orbitals for all cases plus
the OPDOS of the Ca $3p$, Sr $4d$ and Ba $5d$ AOs for the respective
$R$ filled materials. Results obtained for room and high pressure
calculations as presented. Inspecting the figure, one can note that
the components $a=\left(n+1\right)d$ (from either Sr or Ba) and the
$b=3d$ (from Fe) give rise to less OPDOS than the components $a=\left(n+1\right)d$
(from either Sr or Ba) and $b=5p$ (from Sb). Thus, the $(n+1)d+5p$
OPDOS were used to generate the highest superposing highest contributing)
MOs for the filler-cage $\left(n+1\right)d+5p$ bonding. In the $R=$
Ca case, the $a=3p$ (from Ca) and $b=5p$ (from Sb) take part is
the highest contributing MOs. In the main text, we focus our analysis
on these highest contributing MOs, which are plotted as contour plots
in figure $6$ in the main text. 

\begin{figure}[H]
\begin{centering}
\includegraphics[width=1\textwidth]{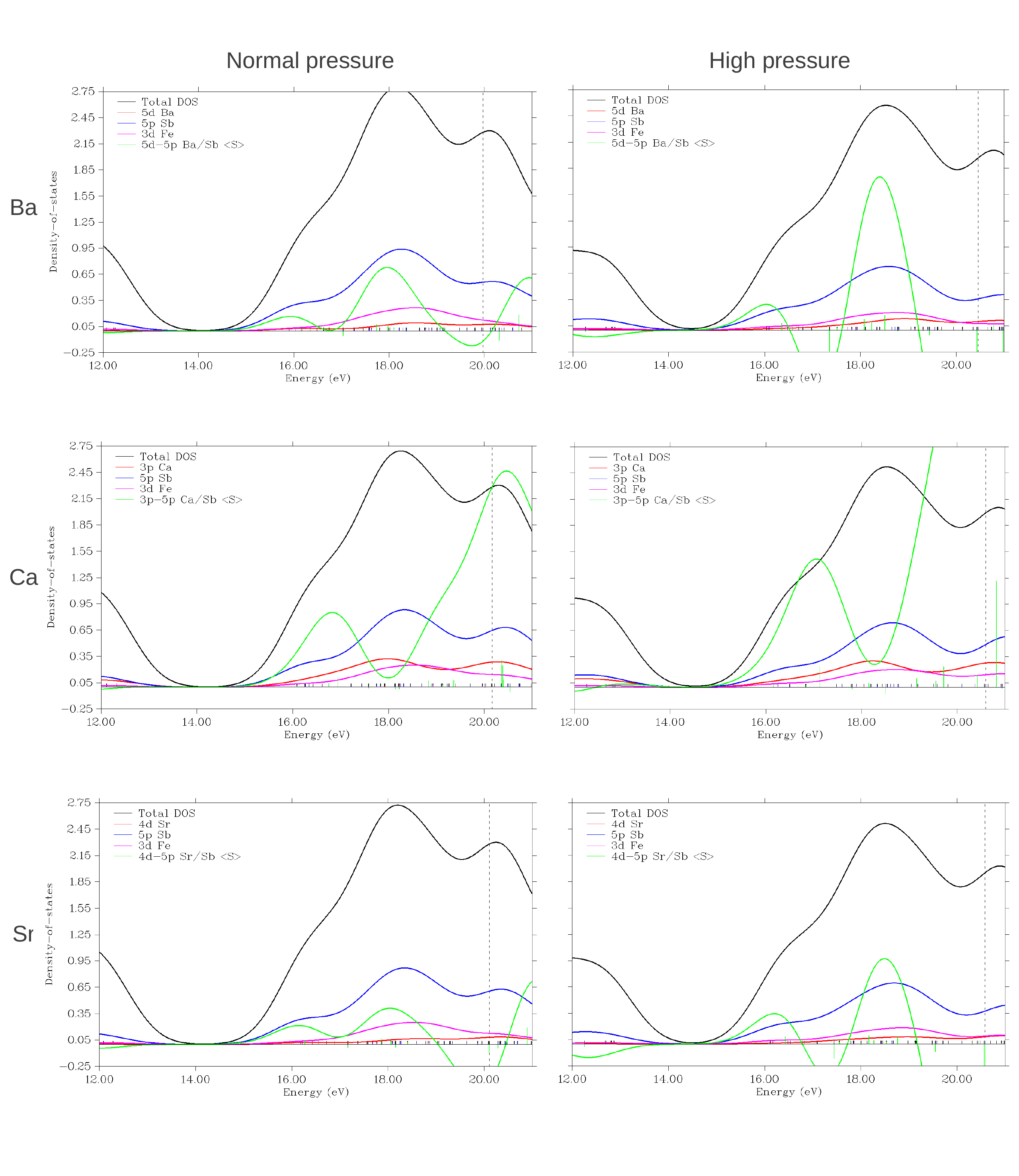}
\par\end{centering}
\caption{Total DOS and OPDOS for the AOs in the vicinity of $E_{F}$ taking
part in the Ca $3p$-$5p$ Sb, Sr $4d$-$5p$ Sb and Ba $5d$-$5p$
Sr bonding MOs. In each panel, the left axis describes the DOS for
all orbitals while the right axis shows the OPDOS. \label{fig:Total-DOS,-partial}}
\end{figure}

\bibliographystyle{apsrev4-2}
\bibliography{2021JuAbrantes_vibrationalElastic_references}